\documentclass[12pt,a4paper]{report}

\usepackage[english]{babel}
\usepackage[latin1]{inputenc}
\usepackage{amsmath}
\usepackage{makeidx}
\usepackage{amssymb}
\usepackage{theorem}
\usepackage{graphicx}
\usepackage{subfigure}
\usepackage{amsfonts}
\usepackage{enumerate}
\usepackage{bbm}
\usepackage{hyperref}

\title{SIRS and games}

\begin{document}

\begin{center}
\Large{\textbf{Nonlinear behavior of coupled Evolutionary Games - Epidemiological Models}}
\end{center}

\begin{center}
Eliza Maria Ferreira$^1$, Ricardo Edem Ferreira$^2$ and Chiara Mocenni$^3$ \\ $^{1, 2}$\textit{Federal University of Lavras} \\ $^{3}$\textit{University of Siena}
\end{center}

\section*{Abstract}

Epidemiological models are an important tool in coping with epidemics, as they offer a forecast, even if often simplistic, of the behavior of the disease in the population. This allows responsible health agencies to organize themselves and adopt strategies to minimize and postpone the population's infection peaks. While during an epidemic outbreak, the available model can be used to describe the behavior of the disease in order to aim for fast and efficient forecasts of the epidemiological scenario, once the epidemiological emergency is over, the objective of the subsequent works is to extend models by integrating new facts and information, thus providing more efficient tools to face future epidemics. In this sense, we present an epidemiological model that takes into account on one hand, the creation of a vaccine during an epidemic outbreak (as we saw happen in the case of COVID-19) and that, on the other hand, considers the impact of the cooperative behavioral choices of individuals.
\
\\
Keywords: Epidemiological Models; Game Theory; Cooperation Game

\section*{Introduction}

Epidemiology is an area of science that studies problems related to the evolution of a disease in the population. In biomathematics, these phenomena can have their dynamics described by mathematical models, which can be very useful to determine the evolution of a possible epidemic and even to determine strategies that interrupt or reduce the spread of the disease, whether through vaccination, quarantine, among others.

The most used models to describe the spread of diseases in the population are the SIR models and their variations. The SIR model was created in 1927 by W. O. Kermack and A. G. McKendrick, \cite{kermack1927contribution}. In the model, the authors considered a constant population of size N divided into three classes: Susceptible, S(t); Infected, I(t); and Removed, R(t). Each of the classes varies according to time, t, and $N = S(t) + I(t) + R(t)$. The susceptible class comprises all individuals who have never acquired the disease. In a population where a certain disease has not yet established itself, all individuals are considered susceptible. The class of infected is composed of individuals who are contaminated by the disease. And finally, the removed class is composed of individuals who have already recovered from the disease and acquired permanent immunity to it. Individuals who are in the removed class are also unable to transmit the disease to susceptible individuals. In this model, the infected individual acquires permanent immunity to the disease. Among the best known variations of the SIR model are: the SI or SIS model, used to describe diseases that do not confer any type of immunity to individuals and the SIRS model, used to describe diseases that confer temporary immunity to individuals, for more details see \cite{britton2005essential, edelstein2005mathematical, murray2007mathematical}.

Other variations of the cited models include classes of latent or asymptomatic individuals, as we can see in \cite{gaeta2020simple, madeo2022identification, neves2020predicting} and classes of vaccinated individuals, for example in \cite{ferreira2020controle}.   

Some recent works, see \cite{amaral2021epidemiological, madeo2022identification}, admit that the rational behavior of individuals can have a relevant influence, for example, on the infection rate of the disease, and make use of game theory to couple decision-making by part of the players (agents) and the epidemiological scenario.

The Game Theory studies situations where two or more players rationally choose their actions in an attempt to maximize their gains or equivalently, minimize their losses. It stood out as a branch of mathematics after the publication of the book \textit{Theory of games and economic behavior} written by John von Neumann and Oskar Morgenstern, \cite{morgenstern1953theory}, and initially focused on solving economic problems. Its results can be applied in different areas, such as biology, philosophy, political science, with applications that can vary from simple entertainment games to relevant aspects of life in society. A well-known example of the application of game theory to social dilemmas is the Prisoner's Dilemma \cite{tucker1983mathematics}. 

In the area of biology, John Maynard Smith and George R. Price called society's attention to the application of game theory to the study of the evolution of some species, with the publication of the article \textit{The logic of animal conflict}, \cite{MaynardSmithPrice}. In Evolutionary Game Theory, whose creation is attributed to John Maynard Smith for formalizing the idea of \textit{evolutionarily stable strategies} in \cite{smith1982evolution}, it is considered a situation (in this case a game) in which individuals of a population of infinite size randomly interact with each other and each interaction generates a ``reward'' for the individuals involved. Originally, this theory was developed to study deterministic dynamics in infinite populations \cite{hofbauersigmund, taylorjonker}.The system of ODEs that quantifies this dynamic, introduced by Taylor and Jonker \cite{taylorjonker} is called the \textit{replicator dynamics}. 

In this work, we propose an epidemiological model that takes into account the decision-making power of individuals. Our model is a variation of the well-known SIRS model, to which we add the possibility of vaccinating individuals, coupled with a cooperation game with three strategies. Regarding the models proposed by \cite{amaral2021epidemiological, madeo2022identification}, our work is different due to the choices of classes of individuals contemplated in the model, the number of cooperation strategies considered in the game and the function responsible for making the connection between the epidemiological scenario and the choice of strategies. We also did not create a model thinking about a specific disease like COVID-19, although the model is suitable for that disease as well. Our goal is to create a generic model that can be adapted to different scenarios later on. 

We started our work by proposing an epidemiological model with vaccination, but without considering the cooperation attitudes of individuals, that is, a model not coupled to any cooperation game. Then we define our cooperation game, not yet coupled to the epidemiological model, to then make the coupling between the epidemiological model and the game. We finish our work with a qualitative study of the dynamics of the coupled model and an application.

\section*{The model SIRS with immunization}

In this section we will use a model that describes the dynamics of a virus (like the COVID-19 virus for example) considering the possibility of reinfection and the existence of a vaccine against the disease. In our model we assume that individuals can receive temporary immunity through the vaccine and that the disease can also generate some temporary immunity to the individual. We consider a dimensionless model with constant population divided into twelve classes: unimmunized susceptible $S_{n}$, susceptible immunized by the disease $S_{i}$, susceptible immunized by vaccine $S_{v}$, infected not immunized $I_n$, infected immunized by the disease $I_{i}$, infected immunized by vaccine $I_v$, recovered not immunized $R_n$, recovered immunized by the disease $R_{i}$, recovered immunized by vaccine $R_v$, dead not immunized $D_n$, dead immunized by the disease $D_n$ and dead immunized by vaccine $D_v$, where $1 = S_n + S_i + S_v + I_n + I_i + I_v + R_n + R_i + R_v + D_n + D_i + D_v$.

\begin{equation}\label{SVIRS game off}
\left\{
\begin{array}{ccl}
\dot{S_{n}} & = & - (\beta_{s}I_{s} + \beta_{i}I_{i} + \beta_{v}I_{v})S_{n} - \phi_{n} S  + \tau_{i} S_{i} + \tau_{v} S_{v} \\
\dot{S_{i}} & = & \alpha_{n} R_{n} + \alpha_{i} R_{i} - \phi_{i} S_{i} - (1-\theta_{i})(\beta_{n}I_{n} + \beta_{i}I_{i} + \beta_{v}I_{v})S_{i} - \tau_{i} S_{i} \\
\dot{S_{v}} & = & \phi_{n} S_{n} + \phi_{i} S_{i} + \alpha_{v} R_{v} - (1-\theta_{v})(\beta_{n}I_{n} + \beta_{i}I_{i} + \beta_{v}I_{v})S_{v} - \tau_{v} S_{v} \\
\dot{I_{n}} & = & (\beta_{s}I_{s} + \beta_{i}I_{i} + \beta_{v}I_{v})S_{n} - \gamma_{n}I_{n} - \lambda_{n}I_{n} \\
\dot{I_{i}} & = & (1-\theta_{i})(\beta_{n}I_{n} + \beta_{i}I_{i} + \beta_{v}I_{v})S_{i} - \gamma_{i}I_{i} - \lambda_{i}I_{i} \\
\dot{I_{v}} & = & (1-\theta_{v})(\beta_{n}I_{n} + \beta_{i}I_{i} + \beta_{v}I_{v})S_{v} - \gamma_{v}I_{v} - \lambda_{v}I_{v} \\
\dot{R_{n}} & = & \gamma_{n}I_{n} - \alpha_{n} R_{n} \\
\dot{R_{i}} & = & \gamma_{i}I_{i} - \alpha_{i} R_{i} \\
\dot{R_{v}} & = & \gamma_{v}I_{v} - \alpha_{v} R_{v} \\
\dot{D_n} & = & \lambda_{n}I_{n} \\
\dot{D_i} & = & \lambda_{i}I_{i} \\
\dot{D_v} & = & \lambda_{v}I_{v}
\end{array}
\right.
\end{equation}

In the model above the term $\beta_{\star}$ indicates the infection rate without immunization $\star = n$, with immunization by disease $\star = i$ and with immunization by vaccine $\star = v$. Analogously, $\gamma_{\star}$ indicates the recuperation rate, $\lambda_{\star}$ the death rate from the disease and $\alpha_{\star}$ the rate at which individuals lose the immunity they have acquired from the disease. The parameter $\phi$ indicates the vaccination rate, $\theta$ the effectiveness of the immunization and $\tau$ the rate at which individuals lose the immunity they have acquired from the disease or vaccine. 

Let $\mathcal{A}$ be the set that contains all the trajectories of the system \ref{SVIRS game off}. This is: 
\begin{equation*}
\mathcal{A} = \left\{y \in \mathbb{R}^{12}_{+} ; \sum_{i=1}^{12} y_i = 1 \right\}.
\end{equation*}
\
\\
Such a system has infinite equilibrium points in $\mathcal{A}$ which can be described as follows:
\begin{eqnarray*}
y^{*}(k_1, k_2, k_3) & = & \left(\dfrac{\tau_{v}}{\phi_{n} + \tau_{v}}(1-k_1-k_2-k_3), 0, \right. \\ & & \left. \dfrac{\phi_{n}}{\phi_{n} + \tau_{v}}(1-k_1-k_2-k_3), 0, 0, 0, 0, 0, 0, k_1, k_2, k_3 \right). 
\end{eqnarray*}

\
\\
Where $k_{1}, k_{2}, k_{3} \in [0, 1]$ and $k_{1} + k_{2} + k_{3} \leq 1$. Note that all equilibriums of dynamic are only reached when there is no more disease in the population, that is, the number of infected with or without immunization is equal to zero.

The model discussed in this section is a variation of the well-known SIRS model widely used to describe diseases that do not generate permanent immunity to the population. These models, however, do not take into account the behavior of individuals in an epidemic situation, such as some type of social isolation, the use of personal protective equipment, among others. We then want to consider some standard behaviors in an epidemic context and create a new model from the \ref{SVIRS game off} model incorporating these behaviors. To incorporate the behavior of individuals into the epidemiological model, we will first define what we will call a ``cooperation game'' to describe the behavior of individuals, the different types of behavior will be seen as the strategies adopted in the game. So let's couple the epidemiological model \ref{SVIRS game off} and the game so that the dynamics of the epidemiological model depends on the strategy chosen in the game and the choice of strategy depends on the epidemiological scenario. Something similar can be seen in \cite{madeo2022identification} and \cite{amaral2021epidemiological}, in the two cited works, only two types of behaviors were considered in the population that were described by a game of two strategies. In this work we are going to include one more type of behavior possible to the model, that is, we are going to couple to the epidemiological model a game with three strategies and to study the dynamics of the new model created.

In the next section, we will see the cooperation game that we will use in this work.

\section*{A cooperation game}
Let us now assume that the population can rationally assume a certain behavior during an epidemic outbreak. Let us consider that individuals in the population can adopt three types of behavior that we will define as strong cooperation $x_1$, weak cooperation $x_2$ and non-cooperation (or defection) $x_3$. We can think of strong cooperation behavior as attitudes that strongly contribute to containing the spread of some disease, such as, for example, the adoption of social isolation together with the use of personal protective equipment. In the weak cooperation behavior, there are actions that have some positive influence on containing the spread of the disease, but in a less effective way compared to strong cooperation. An example of weak cooperation can be the use of personal protective equipment without adopting social isolation. And we can think of non-cooperative behavior as actions that do not or contribute negatively to disease containment. An example of non-cooperation can be the normal routine of each individual without adopting any practice to contain a particular disease.

Next, we propose a game between the agents involved considering the mentioned behaviors as the pure strategies of the game, where $1 = x_1 + x_2 + x_3$. The pay-off matrix is 

\begin{equation}\label{pay-off matrix}
M=\begin{pmatrix}
1 & A & B \\ C & \frac{1}{2} & E \\ F & G & 0
\end{pmatrix}.
\end{equation}

In the matrix above, the element of row i and column j indicates the reward that the individual who plays the pure strategy $x_i$ receives when interacting with the individual who plays the pure strategy $x_j$. 

Note that we are considering that the reward received by an individual when interacting with another that adopts the same strategy is fixed. It is also reasonable to expect that the reward received by an individual, regardless of the strategy chosen by him, when interacting with an individual who adopts the strong cooperation strategy is greater than the reward received when interacting with an individual who adopts the weak cooperation strategy, which in turn is greater than the reward received when interacting with an individual who adopts the non-cooperation strategy. For this reason we will consider that $1 > A > B$, $C > \frac{1}{2} > E$ and $F > G > 0$ in the pay-off matrix \ref{pay-off matrix}.

Here we will introduce some useful functions that will highlight the relationship between the parameters of the payment matrix M. We will use these functions to define what we will call cooperation scenarios. Let

\begin{equation}\label{sigma0}
\begin{array}{ccl}
\sigma_{x_1x_2} & = & 1 - C\\
\sigma_{x_2x_1} & = & \frac{1}{2} - A\\
\sigma_{x_1x_3} & = & 1 - F\\
\sigma_{x_3x_1} & = & - B\\
\sigma_{x_2x_3} & = &\frac{1}{2} - G\\
\sigma_{x_3x_2} & = & - E.
\end{array}
\end{equation}

Note that for a fixed payoff matrix the $\sigma$ , defined in \ref{sigma0}, are constants. When $\sigma_{x_1x_2}>0$, $\sigma_{x_2x_1}<0$, $\sigma_{x_1x_3}>0$, $\sigma_{x_3x_1}<0$, $\sigma_{x_2x_3}>0$ and $\sigma_{x_3x_2}<0$ we have what we call a strong cooperation scenario. And we have two possibilities for a weak cooperation scenario, $\sigma_{x_1x_2}<0$, $\sigma_{x_2x_1}>0$, $\sigma_{x_2x_3}>0$, $\sigma_{x_3x_2}<0$, $\sigma_{x_1x_3}>0$, $\sigma_{x_3x_1}<0$ and $\sigma_{x_1x_2}<0$, $\sigma_{x_2x_1}>0$, $\sigma_{x_2x_3}>0$, $\sigma_{x_3x_2}<0$, $\sigma_{x_1x_3}<0$, $\sigma_{x_3x_1}>0$. In the first case non-collaboration is the worst strategy, in the second case the worst strategy is strong collaboration. When $\sigma_{x_1x_2}<0$, $\sigma_{x_2x_1}>0$, $\sigma_{x_1x_3}<0$, $\sigma_{x_3x_1}>0$, $\sigma_{x_2x_3}<0$ and $\sigma_{x_3x_2}>0$ we have a scenario of non-collaboration.

We will study the cooperation game that we defined in this section from a deterministic point of view. Thus, the dynamics of the game will be given, as we will see later, by the dynamics of the replicator, introduced by Taylor and Jonker \cite{taylorjonker}.

\subsection*{The replicator dynamics}
Let's take a look at the dynamics of what we're calling a ``cooperation game''. The variation of players who opt for the $x_1$, $x_2$ or $x_3$ strategies can be calculated by replicator dynamics:

\begin{equation}\label{cooperation game}
\left\{\begin{array}{ccl}
\dot{x_1} & = & x_1(\pi_{x_1} - \bar{\pi}) \\
\dot{x_2} & = & x_2(\pi_{x_2} - \bar{\pi}) \\
\dot{x_3} & = & x_3(\pi_{x_3} - \bar{\pi}) 
\end{array}\right.
\end{equation}
\
\\
where,
$$\left[\begin{array}{c} \pi_{x_1} \\ \pi_{x_2} \\ \pi_{x_3} \end{array}\right] = M\left[\begin{array}{c} x_1 \\ x_2 \\ x_3 \end{array}\right] = \left[\begin{array}{c} x_1 + Ax_2 + Bx_3 \\ Cx_1 + \frac{1}{2}x_2 + Ex_3  \\ Fx_1 + Gx_2 \end{array}\right]$$
\
\\
and $\bar{\pi} = x_1\pi_{x_1} + x_2\pi_{x_2} + x_3\pi_{x_3}$. The $\pi_{x_i}$ functions indicate the average payoff of the $x_i$ strategy and $\bar{\pi}$ is the total average payoff.  
Writing in terms of the $\sigma$ functions, \ref{sigma0}, we have:

\begin{eqnarray}
\pi_{x_1} & = & x_1 + (\frac{1}{2} - \sigma_{x_2x_1})x_2 - \sigma_{x_3x_1}(1-x_1-x_2)  \label{pi x1}\\
\pi_{x_2} & = & (1 - \sigma_{x_1x_2})x_1 + \frac{1}{2}x_2 - \sigma_{x_3x_2}(1-x_1-x_2) \label{pi x2}\\
\pi_{x_3} & = & (1 - \sigma_{x_1x_3})x_1 + (\frac{1}{2}- \sigma_{x_2x_3})x_2 \label{pi x3}\\
\bar{\pi} & = & x_1^2 + (1 - \sigma_{x_1x_3} - \sigma_{x_3x_1})x_1(1-x_1) \nonumber\\ & & + \frac{1}{2}x_2
^2 + (\frac{1}{2} - \sigma_{x_2x_3} - \sigma_{x_3x_2})x_2(1-x_2) \nonumber\\ & & + (- \sigma_{x_2x_1} - \sigma_{x_1x_2} + \sigma_{x_3x_1} + \sigma_{x_3x_2} + \sigma_{x_1x_3} + \sigma_{x_2x_3})x_1x_2 \label{bar pi}
\end{eqnarray}

See that the average payoff of one strategy is better or worse than the average payoff of another strategy depending on the frequency $x_i$ of each strategy and also on the sign of the $\sigma$ functions.  

So far, we have seen how the epidemiological dynamics and the cooperation game dynamics given by the replicator dynamics work separately. Next, we will see how to couple the cooperation game to the epidemiological dynamics and vice versa.  

\section*{Coupling the game and the epidemiological model}

We can think that during the progression of a given disease, agents can rationally choose between strong cooperation, weak cooperation and non-cooperation behaviors. So we are assuming that the agents of the epidemiological model are playing the cooperation game simultaneously. Obviously, data related to the epidemic can influence the choice of a certain behavior by an agent and the behavior of agents can also influence the dynamics of the epidemiological model.

Let's see below how data related to the epidemic can influence the choice of a certain behavior or cooperation strategy.

Let $a$, $b$ and $c$ be real numbers such that $0 < c < b < a < 1$. And let $w(t) = S_{v}(t) + I_v(t) + R_v(t) + D_v(t)$ be the total fraction of the vaccinated population at time $t$. We define the functions

\begin{equation}\label{G functions}
\left\{\begin{array}{rcl} G_a(I,\dot{D},w) & = & (1-hw)I + \dot{D} - a \\G_b(I,\dot{D},w) & = & (1-hw)I + \dot{D} - b \\G_c(I,\dot{D},w) & = & (1-hw)I + \dot{D} - c \end{array}\right.
\end{equation}

\
\\
where $I = I_n + I_i + I_v$, $\dot{D} = \dot{D_n} + \dot{D_i} + \dot{D_v}$ and $h \in [0, 1]$ is a parameter that quantifies the influence of information about the total fraction of vaccinated people on the population's concerns with the total amount of infected people. For example, for $h$ close to 1 and relatively high $w(t)$ the total number of infected at the instant $t$, $I(t)$, has less relevance compared to the situation where $w(t)$ is small because the symptoms of the disease in question in individuals who have been vaccinated are milder and not of great concern. For $h$ close to zero, the total fraction of vaccinated individuals has no influence on the concern of individuals with the total number of infected.

We have already seen that the choice of the best strategy (the one that maximizes the gains in the game) is associated with the signs of the $\sigma$ functions. Let's assume that we initially have a scenario of strong cooperation (strong cooperation is the best strategy) i.e. $\sigma^{0}_{x_1x_2}>0$, $\sigma^{0}_{x_2x_1}<0$, $\sigma^{0}_{x_1x_3}>0$, $\sigma^{0}_{x_3x_1}<0$, $\sigma^{0}_{x_2x_3}>0$ and $\sigma^{0}_{x_3x_2}<0$. So we define

\begin{equation}\label{sigma}
\begin{array}{ccl}
\sigma_{x_1x_2} & = & \sigma^0_{x_1x_2}G_a(I,\dot{D},w)\\
\sigma_{x_2x_1} & = & \sigma^0_{x_2x_1}G_a(I,\dot{D},w)\\
\sigma_{x_1x_3} & = & \sigma^0_{x_1x_3}G_b(I,\dot{D},w)\\
\sigma_{x_3x_1} & = & \sigma^0_{x_3x_1}G_b(I,\dot{D},w)\\
\sigma_{x_2x_3} & = & \sigma^0_{x_2x_3}G_c(I,\dot{D},w)\\
\sigma_{x_3x_2} & = & \sigma^0_{x_3x_2}G_c(I,\dot{D},w)
\end{array}
\end{equation}

Unlike the $\sigma$ functions defined in \ref{sigma0}, here the $\sigma$ functions are not constant, they depend on the functions defined in \ref{G functions}, ie they depend on the epidemiological scenario.

Using \ref{sigma} and the relations obtained in \ref{pi x1}, \ref{pi x2}, \ref{pi x3} and \ref{bar pi} we can rewrite \ref{cooperation game} as follows:

\begin{equation}\label{cooperation game sigma}
\left\{\begin{array}{rl}
\dot{x_1} = & x_1\left[x_2((\sigma^0_{x_1x_2} + \sigma^0_{x_2x_1})x_1 - \sigma^0_{x_2x_1})G_{a}(I,\dot{D},w)\right. \\ 
& \left. + x_3((\sigma^0_{x_1x_3} + \sigma^0_{x_3x_1})x_1 - \sigma^0_{x_3x_1})G_{b}(I,\dot{D},w)\right. \\
& \left. + x_3x_2(\sigma^0_{x_2x_3} + \sigma^0_{x_3x_2})G_{c}(I,\dot{D},w)\right]\\
\dot{x_2} = & x_2\left[x_1((\sigma^0_{x_1x_2} + \sigma^0_{x_2x_1})x_2 - \sigma^0_{x_1x_2})G_{a}(I,\dot{D},w)\right. \\ 
& \left. + x_3x_1(\sigma^0_{x_1x_3} + \sigma^0_{x_3x_1})G_{b}(I,\dot{D},w)\right. \\
& \left. + x_3((\sigma^0_{x_2x_3} + \sigma^0_{x_3x_2})x_2 - \sigma^0_{x_3x_2})G_{c}(I,\dot{D},w)\right].\\
\dot{x_3} = & x_3\left[x_1x_2(\sigma^0_{x_1x_2} + \sigma^0_{x_2x_1})G_{a}(I,\dot{D},w)\right. \\ 
& \left. + x_1((\sigma^0_{x_1x_3} + \sigma^0_{x_3x_1})x_3 - \sigma^0_{x_1x_3})G_{b}(I,\dot{D},w)\right. \\
& \left. + x_2((\sigma^0_{x_2x_3} + \sigma^0_{x_3x_2})x_3 - \sigma^0_{x_2x_3})G_{c}(I,\dot{D},w)\right]
\end{array}\right.
\end{equation}
\
\\
Remembering that $x_3 = 1-x_1-x_2$.

As we have seen, the $G_{\star}$ functions defined in \ref{G functions}, make the choice of a strategy in the game depend on the states of the epidemiological model. We will now define how the choice of a certain strategy in the game can influence the epidemiological dynamics. This relationship between the game and the epidemiological model will be made explicit by defining an infection rate that depends on the game's strategies as follows:

\begin{equation}\label{infection rates}
\left\{\begin{array}{ccl}
\beta_n(x_1, x_2) & = & \beta_{n_0}(1 - e_1x_1 - e_2x_2)\\
\beta_i(x_1, x_2) & = & \beta_{i_0}(1 - e_1x_1 - e_2x_2)\\
\beta_v(x_1, x_2) & = & \beta_{v_0}(1 - e_1x_1 - e_2x_2)
\end{array}\right.
\end{equation}
\
\\
where $1 \geq e_1 > e_2 \geq 0$ and $\beta_{n_0}$, $\beta_{i_0}$, $\beta_{v_0}$ are the infection rates when no cooperation is considered, i.e. all of the population using the $x_3$ strategy. See which strong or weak cooperation behaviors can lower the infection rate. And it is reasonable to think that strong cooperative behavior plays a more effective role in decreasing the infection rate, so we do $e_1 > e_2$.

Next, we will consider the proposed epidemiological model coupled with a cooperation game and make a quantitative analysis of its equilibrium points.

\subsection*{The model equilibrium points}
Adding the replicator dynamics equations to the epidemiological model \ref{SVIRS game off} with the appropriate adjustments that relate to each other, we arrive at the following model:

\begin{equation}\label{SVIRS game on}
\left\{
\begin{array}{ccl}
\dot{S_{n}} & = & - (\beta_{s}(x_1, x_2)I_{s} + \beta_{i}(x_1, x_2)I_{i} + \beta_{v}(x_1, x_2)I_{v})S_{n} - \phi_{n} S \\ & & + \tau_{i} S_{i} + \tau_{v} S_{v} \\
\dot{S_{i}} & = & \alpha_{n} R_{n} + \alpha_{i} R_{i} - \phi_{i} S_{i} - (1-\theta_{i})(\beta_{n}(x_1, x_2)I_{n} \\ & & + \beta_{i}(x_1, x_2)I_{i} + \beta_{v}(x_1, x_2)I_{v})S_{i} - \tau_{i} S_{i} \\
\dot{S_{v}} & = & \phi_{n} S_{n} + \phi_{i} S_{i} + \alpha_{v} R_{v} - (1-\theta_{v})(\beta_{n}(x_1, x_2)I_{n} \\ & & + \beta_{i}(x_1, x_2)I_{i} + \beta_{v}(x_1, x_2)I_{v})S_{v} - \tau_{v} S_{v} \\
\dot{I_{n}} & = & (\beta_{s}(x_1, x_2)I_{s} + \beta_{i}(x_1, x_2)I_{i} + \beta_{v}(x_1, x_2)I_{v})S_{n} - \gamma_{n}I_{n} \\ & & - \lambda_{n}I_{n} \\
\dot{I_{i}} & = & (1-\theta_{i})(\beta_{n}(x_1, x_2)I_{n} + \beta_{i}(x_1, x_2)I_{i} + \beta_{v}(x_1, x_2)I_{v})S_{i} \\ & & - \gamma_{i}I_{i} - \lambda_{i}I_{i} \\
\dot{I_{v}} & = & (1-\theta_{v})(\beta_{n}(x_1, x_2)I_{n} + \beta_{i}(x_1, x_2)I_{i} + \beta_{v}(x_1, x_2)I_{v})S_{v} \\ & & - \gamma_{v}I_{v} - \lambda_{v}I_{v} \\
\dot{R_{n}} & = & \gamma_{n}I_{n} - \alpha_{n} R_{n} \\
\dot{R_{i}} & = & \gamma_{i}I_{i} - \alpha_{i} R_{i} \\
\dot{R_{v}} & = & \gamma_{v}I_{v} - \alpha_{v} R_{v} \\
\dot{D_n} & = & \lambda_{n}I_{n} \\
\dot{D_i} & = & \lambda_{i}I_{i} \\
\dot{D_v} & = & \lambda_{v}I_{v} \\
\dot{x_1} & = & x_1\left[x_2((\sigma^0_{x_1x_2} + \sigma^0_{x_2x_1})x_1 - \sigma^0_{x_2x_1})G_{a}(I,\dot{D},w)\right. \\ 
&& \left. + (1-x_1-x_2)((\sigma^0_{x_1x_3} + \sigma^0_{x_3x_1})x_1 - \sigma^0_{x_3x_1})G_{b}(I,\dot{D},w)\right. \\
&& \left. + (1-x_1-x_2)x_2(\sigma^0_{x_2x_3} + \sigma^0_{x_3x_2})G_{c}(I,\dot{D},w)\right]\\
\dot{x_2} & = & x_2\left[x_1((\sigma^0_{x_1x_2} + \sigma^0_{x_2x_1})x_2 - \sigma^0_{x_1x_2})G_{a}(I,\dot{D},w)\right. \\ 
&& \left. + (1-x_1-x_2)x_1(\sigma^0_{x_1x_3} + \sigma^0_{x_3x_1})G_{b}(I,\dot{D},w)\right. \\
&& \left. + (1-x_1-x_2)((\sigma^0_{x_2x_3} + \sigma^0_{x_3x_2})x_2 - \sigma^0_{x_3x_2})G_{c}(I,\dot{D},w)\right].
\end{array}
\right.
\end{equation}

All trajectories of the \ref{SVIRS game on} system transit in the set
\begin{eqnarray*}
\mathcal{F} & = & \left\{y \in \mathbb{R}^{14}_{+} ; \sum_{i=1}^{12} y_1 = 1,  y_{13} + y_{14}  \leq 1 \right\}.
\end{eqnarray*}

The above system \ref{SVIRS game on} has infinite equilibrium points in $\mathcal{F}$ which can be described as follows:

\begin{eqnarray*}
y^{*}_1(k_1, k_2, k_3) & = & \left(\dfrac{\tau_{v}}{\phi_{n} + \tau_{v}}(1-k_1-k_2-k_3), 0, \right.\\ & & \left. \dfrac{\phi_{n}}{\phi_{n} + \tau_{v}}(1-k_1-k_2-k_3), 0, 0, 0, 0, 0, 0, k_1, k_2, k_3, 1, 0 \right) \\
y^{*}_2(k_4, k_5, k_6) & = & \left(\dfrac{\tau_{v}}{\phi_{n} + \tau_{v}}(1-k_1-k_2-k_3), 0, \right.\\ & & \left. \dfrac{\phi_{n}}{\phi_{n} + \tau_{v}}(1-k_1-k_2-k_3), 0, 0, 0, 0, 0, 0, k_4, k_5, k_6, 0, 1 \right) \\
y^{*}_3(k_7, k_8, k_9) & = & \left(\dfrac{\tau_{v}}{\phi_{n} + \tau_{v}}(1-k_1-k_2-k_3), 0, \right.\\ & & \left. \dfrac{\phi_{n}}{\phi_{n} + \tau_{v}}(1-k_1-k_2-k_3), 0, 0, 0, 0, 0, 0, k_7, k_8, k_9, 0, 0 \right).
\end{eqnarray*}

\
\\
Where $k_{3i-2}, k_{3i-1}, k_{3i} \in [0, 1]$ and $k_{3i-2} + k_{3i-1} + k_{3i} \leq 1$ for all $i = 1, 2, 3$.

We are going to study the behavior of equilibrium points and for that we will analyze the behavior of the trajectories of the \ref{SVIRS game on} system taking initial points sufficiently close to the equilibrium. The main differences between the equilibrium points $y^{*}_1(k_1, k_2, k_3)$, $y^{*}_2(k_4, k_5, k_6)$ and $y^{*}_3(k_7, k_8, k_9)$ are in the last two coordinates: in $y^{*}_1$ equilibrium is reached when the population is composed only of strongly collaborating individuals, in $y^{*}_2$ by weakly collaborating individuals, and in $y^{*}_3 $ equilibrium is reached when the population is composed only of non-collaborating individuals. So let's focus on the behavior of trajectories with respect to these last two coordinates. 

All equilibrium points are reached when we have $I_n = I_i = I_v = 0$ and $\dot{D_n} = \dot{D_i} = \dot{D_v} = 0$. We can consider initial conditions close to the equilibrium points with $I_n$, $I_i$, $I_v$, $\dot{D_n}$, $\dot{D_i}$ and $\dot{D_v}$ as close to zero as we want, so that $G_{c}( I,\dot{D},w) < 0$ and consequently $G_{b}(I,\dot{D},w) < 0$ and $G_{a}(I,\dot{D},w ) < 0$ because $a > b > c \Rightarrow G_c > G_b > G_a$.

We'll start by looking at initial conditions close to $y^{*}_{1}(k_1, k_2, k_3)$. As mentioned earlier, at points close enough to the equilibrium points we have $0 > G_c > G_b > G_a$. Additionally at points close to $y^{*}_1(k_1, k_2, k_3)$, $x_1$ is close enough to $1$ (with values less than $1$) and $x_2$ and $x_3$ are close enough to zero ( with values greater than zero). Therefore, the terms that exert the greatest influence on the sign of $\dot{x_1}$ (see \ref{cooperation game sigma}) are:

\begin{center}
$x_1[x_2((\sigma^0_{x_1x_2} + \sigma^0_{x_2x_1})x_1 - \sigma^0_{x_2x_1})G_{a}(I,\dot{D},w)]$
\end{center} 
and
\begin{center} 
$x_1[x_3((\sigma^0_{x_1x_3} + \sigma^0_{x_3x_1})x_1 - \sigma^0_{x_3x_1})G_{b}(I,\dot{D},w)]$.
\end{center} 
\
\\
The term $x_1[x_3x_2(\sigma^0_{x_2x_3} + \sigma^0_{x_3x_2})G_{c}(I,\dot{D},w)]$ which also appears in $\dot{x_1} $, is multiplied by $x_1$ and $x_2$, two factors that are as close to zero as you like. Unlike the other two terms mentioned that appear multiplied by just a value close to zero. So this term doesn't have much influence on the sign of $\dot{x_1}$. Notice that,

\begin{center}
$((\sigma^0_{x_1x_2} + \sigma^0_{x_2x_1})x_1 - \sigma^0_{x_2x_1}) = (x_1\sigma^0_{x_1x_2} + \sigma^0_{x_2x_1}(x_1 - 1)) > 0$
\end{center}
and
\begin{center}
$((\sigma^0_{x_1x_3} + \sigma^0_{x_3x_1})x_1 - \sigma^0_{x_3x_1}) = (x_1\sigma^0_{x_1x_3} + \sigma^0_{x_3x_1}(x_1 - 1)) > 0$.
\end{center}
\
\\
So at points close enough to $y^{*}_1(k_1, k_2, k_3)$, $\dot{x_1}<0$.

The term with the greatest influence on the sign of $\dot{x_2}$ is
\begin{center}
$x_2[x_1((\sigma^0_{x_1x_2} + \sigma^0_{x_2x_1})x_2 - \sigma^0_{x_1x_2})G_{a}(I,\dot{D},w)]$
\end{center}
\
\\
which is positive.

And 
\begin{center}
$x_3[x_1((\sigma^0_{x_1x_3} + \sigma^0_{x_3x_1})x_3 - \sigma^0_{x_1x_3})G_{b}(I,\dot{D},w)]>0$
\end{center} 
\
\\
is the term with the greatest influence on the sign of $\dot{x_3}$. Note that
\begin{center}
$((\sigma^0_{x_1x_3} + \sigma^0_{x_3x_1})x_3 - \sigma^0_{x_1x_3}) = (x_3\sigma^0_{x_3x_1} + \sigma^0_{x_1x_3}(x_3 - 1))<0$.
\end{center} 

So, at points close enough to $y^{*}_1(k_1, k_2, k_3)$, $\dot{x_1}<0$, $\dot{x_2}>0$ and $\dot{x_3}>0 $. That is, the orbits are moving away from $y^{*}_1(k_1, k_2, k_3)$ for all $k_1$, $k_2$, $k_3$, which suggests that these equilibrium are unstable.

Let's now analyze initial conditions close to $y^{*}_2(k_4, k_5, k_6)$.

At points close to $y^{*}_2(k_4, k_5, k_6)$, $x_2$ is close enough to $1$ (with values less than $1$) and $x_1$ and $x_3$ are close enough to zero (with values greater than zero). Similar to the analysis done in the previous case, let's look at the most influential term in each of the derivatives in \ref{cooperation game sigma}.

For points close enough to $y^{*}_2(k_4, k_5, k_6)$, the term that most influences the sign of $\dot{x_1}$ is
\begin{center}
$x_1[x_2((\sigma^0_{x_1x_2} + \sigma^0_{x_2x_1})x_1 - \sigma^0_{x_2x_1})G_{a}(I,\dot{D},w)]<0$.
\end{center}

The term that most influences the sign of $\dot{x_3}$ is
\begin{center}
$x_3[x_2((\sigma^0_{x_2x_3} + \sigma^0_{x_3x_2})x_3 - \sigma^0_{x_2x_3})G_{c}(I,\dot{D},w)]>0$.
\end{center}
\
\\
Then $\dot{x_1}<0$ e $\dot{x_3}>0$.

The sign of $\dot{x_2}$ is influenced by the terms
\begin{center}
$x_2[x_1((\sigma^0_{x_1x_2} + \sigma^0_{x_2x_1})x_2 - \sigma^0_{x_1x_2})G_{a}(I,\dot{D},w)]>0$
\end{center}
\
\\
and
\begin{center}
$x_2[x_3((\sigma^0_{x_2x_3} + \sigma^0_{x_3x_2})x_2 - \sigma^0_{x_3x_2})G_{c}(I,\dot{D},w)]<0$.
\end{center}
\
\\
But we have already seen that at points close to $y^{*}_2(k_4, k_5, k_6)$, $\dot{x_1}<0$ and $\dot{x_3}>0$, so the term $x_2[x_3( (\sigma^0_{x_2x_3} + \sigma^0_{x_3x_2})x_2 - \sigma^0_{x_3x_2})G_{c}(I,\dot{D},w)]$ exerts greater influence on the sign of $\dot{x_2}$ compared to the term $x_2[x_1((\sigma^0_{x_1x_2} + \sigma^0_{x_2x_1})x_2 - \sigma^0_{x_1x_2})G_{a}(I,\dot{D},w)]$.

So, at points close enough to $y^{*}_2(k_4, k_5, k_6)$, $\dot{x_1}<0$, $\dot{x_2}<0$ and $\dot{x_3}> 0$ for all $k_4$, $k_5$, $k_6$. Which suggests that these balances are saddle-like.

It now remains to analyze the initial conditions close to $y^{*}_3(k_7, k_8, k_9)$.

At points close to $y^{*}_3(k_7, k_8, k_9)$, $x_3$ is close enough to $1$ (with values less than $1$) and $x_1$ and $x_2$ are close enough to zero (with values greater than zero).

For points close enough to $y^{*}_2(k_7, k_8, k_9)$, the term that most influences the sign of $\dot{x_1}$ is
\begin{center}
$x_1[x_2((\sigma^0_{x_1x_3} + \sigma^0_{x_3x_1})x_1 - \sigma^0_{x_3x_1})G_{a}(I,\dot{D},w)]<0$.
\end{center}

The term that most influences the sign of $\dot{x_2}$ is 
\begin{center}
$x_2[x_3((\sigma^0_{x_2x_3} + \sigma^0_{x_3x_2})x_2 - \sigma^0_{x_3x_2})G_{c}(I,\dot{D},w)]<0$.
\end{center} 

The sign of $\dot{x_3}$ is influenced by the terms
\begin{center}
$x_3[x_1((\sigma^0_{x_1x_3} + \sigma^0_{x_3x_1})x_3 - \sigma^0_{x_1x_3})G_{b}(I,\dot{D},w)]$
\end{center}
\
\\
and
\begin{center}
$x_3[x_2((\sigma^0_{x_2x_3} + \sigma^0_{x_3x_2})x_3 - \sigma^0_{x_2x_3})G_{c}(I,\dot{D},w)],$
\end{center}
\
\\
both positive.

So, at points close enough to $y^{*}_3(k_7, k_8, k_9)$, $\dot{x_1}<0$, $\dot{x_2}<0$ and $\dot{x_3}>0 $ for all $k_7$, $k_8$, $k_9$. Which suggests that these equilibrium are stable.

The result of this analysis is quite interesting, since when the equilibrium points are reached it means that there are no more infected people in the population, so it does not make sense to maintain cooperative behaviors, which is in accordance with the analysis made, where the ``stable equilibrium'' are the equilibrium of the type $y^{*}_3(k_7, k_8, k_9)$ where there is no cooperation. 

\subsection*{An aplication}
In this section we will present a possible application for the proposed model. We will use the model presented in \ref{SVIRS game on} to describe the dynamics of a virus in the population, such as Covid-19 for example, taking into account the behavior of the population, that is, taking into account attitudes of strong cooperation, weak cooperation and non-cooperation. Note that the model presented in \ref{SVIRS game on} is dimensionless, that is, we consider the fraction of individuals in each compartment. However, in the simulations that we carried out and will present below, to reduce computational approximation errors, instead of the fraction of individuals, we considered the number of individuals in each compartment for a population of any size N. The parameters used are random, that is, they are not related to any specific disease. 

We simulate the arrival of a virus in an entirely susceptible population and initially without a vaccine for the disease in question. And later, with the data obtained in the previous simulations, we simulated the dynamics of the disease in a population already affected by the disease in question but with a vaccination process in progress. As with Covid-19, we think of a population capable of creating a vaccine against a given pathogen before the end of an epidemic caused by that pathogen.

\begin{figure}[h]\label{fig1}
	\begin{center}
	\subfigure[Infected variation graph. \label{infectadosS}]{
		\includegraphics[width= 0.4\textwidth]{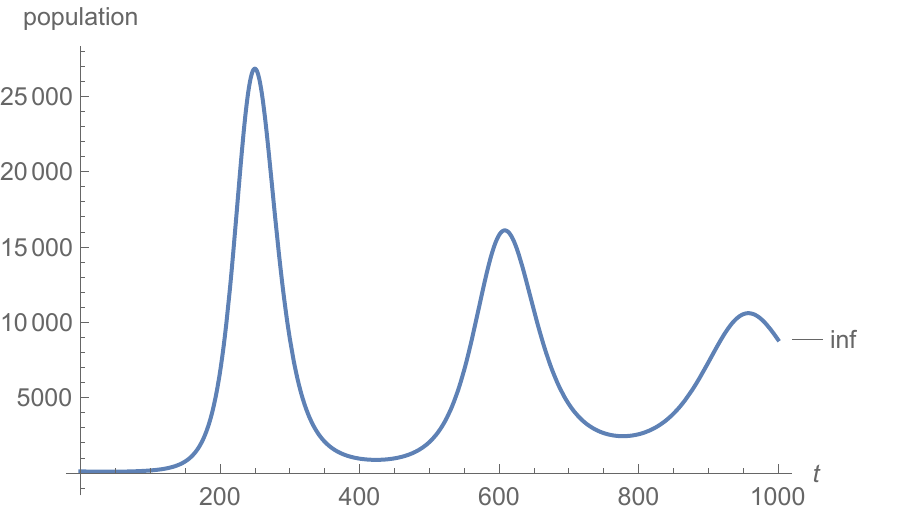}
		}
			\quad
	\subfigure[Graph of variation of behaviors. \label{comportamentosS}]{
	  \includegraphics[width= 0.4\textwidth]{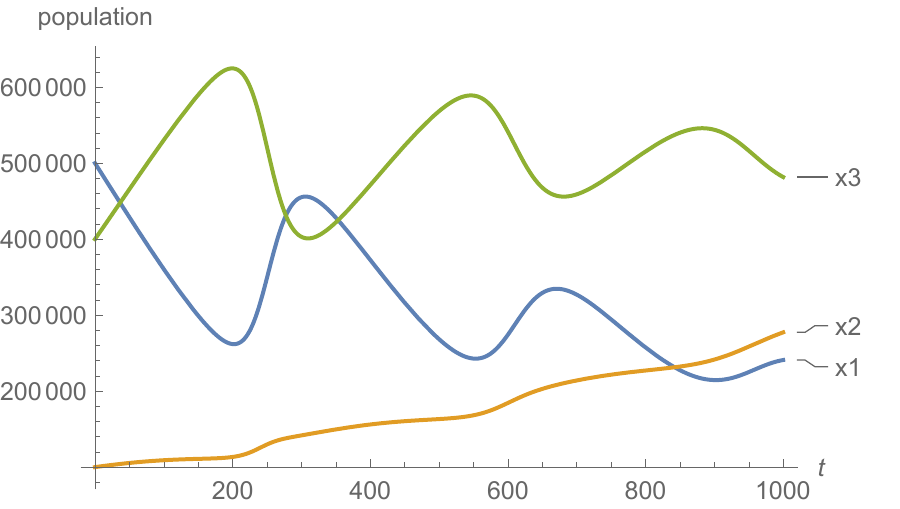}
		}		
	\end{center}
\caption{Infected and collaboration graph without vaccination process. The parameter values used in these simulations were: $\beta_{n_0} = 0.29$, $\beta_{i_0} = 0.22$, $\beta_{v_0} = 0.22$, $\phi_n = 0$, $\phi_i = 0$, $\theta_i = 0.01$, $\theta_v = 0.01$, $\tau_i = 0.003$, $\tau_v = 0.003$, $\gamma_{n} = 1/7$, $\gamma_{i} = 1/5$, $\gamma_{v} = 1/5$, $\lambda_{n} = 0.0028$, $\lambda_{i} = 0.00000002$, $\lambda_{v} = 0.00000002$, $\alpha_n = 1/30$, $\alpha_i = 1/30$, $\alpha_v = 1/30$, $e_1 = 0.95$, $e_2 = 0.7$, $h = 0.9$, $a = 0.01$, $b = 0.007$, $c = 0.004$, pay-off matrix $\mbox{M} = \left[\begin{array}{ccc} 1 & 0.92 & 0.87 \\ 0.59 & 0.5 & 0.46 \\ 0.1 & 0.03 & 0\end{array}\right]$ and initial conditions $S_n(0) = 0.9999 N$, $S_i(0) = 0$, $S_v(0) = 0$, $I_n(0) = 0.0001 N$, $I_i(0) = 0$, $I_v(0) = 0$, $R_n(0) = 0$, $R_i(0) = 0$, $R_v(0) = 0$, $D_n(0) = 0$, $D_i(0) = 0$, $D_v(0) = 0$, $x_1(0) = 0.5 N$ and $x_2(0) = 0.1 N$ where $N = 1000000$.
 }
\end{figure}

See in \ref{infectadosS} and \ref{comportamentosS} the total variation of infected over time and the variation of strong, weak and uncooperative behaviors in the population. Considering that in this population there is still no vaccine to control the spread and effects of the disease.

\begin{figure}[h]\label{fig2}
	\begin{center}
	\subfigure[Infected variation graph. \label{infectadosV}]{
		\includegraphics[width= 0.4\textwidth]{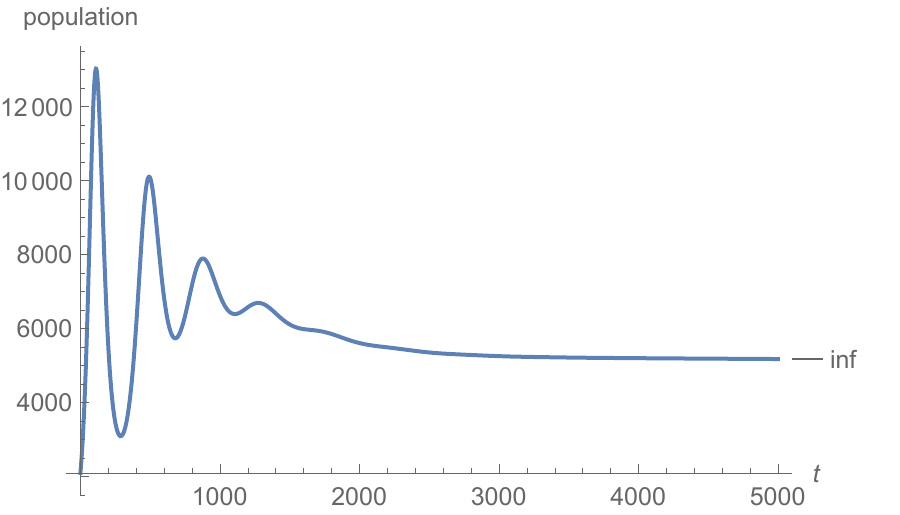}
		}
			\quad
	\subfigure[Graph of variation of behaviors. \label{comportamentosV}]{
	  \includegraphics[width= 0.4\textwidth]{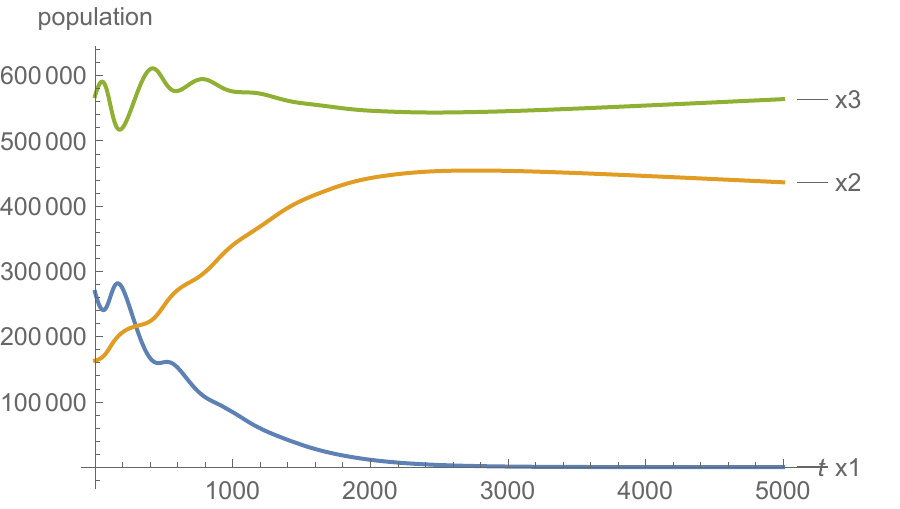}
		}		
	\end{center}
\caption{Infected and collaboration graph with vaccination process. The parameter values used in these simulations were: $\beta_{n_0} = 0.29$, $\beta_{i_0} = 0.22$, $\beta_{v_0} = 0.22$, $\phi_n = 0.0016$, $\phi_i = 0$, $\theta_i = 0.01$, $\theta_v = 0.01$, $\tau_i = 0.003$, $\tau_v = 0.003$, $\gamma_{n} = 1/7$, $\gamma_{i} = 1/5$, $\gamma_{v} = 1/5$, $\lambda_{n} = 0.0028$, $\lambda_{i} = 0.00000002$, $\lambda_{v} = 0.00000002$, $\alpha_n = 1/30$, $\alpha_i = 1/30$, $\alpha_v = 1/30$, $e_1 = 0.95$, $e_2 = 0.7$, $h = 0.9$, $a = 0.01$, $b = 0.007$, $c = 0.004$, pay-off matrix $\mbox{M} = \left[\begin{array}{ccc} 1 & 0.92 & 0.87 \\ 0.59 & 0.5 & 0.46 \\ 0.1 & 0.03 & 0\end{array}\right]$ and initial conditions $S_n(0) = 0.8163 N$, $S_i(0) = 0.1689$, $S_v(0) = 0$, $I_n(0) = 0.0018 N$, $I_i(0) = 0.0003$, $I_v(0) = 0$, $R_n(0) = 0.0053$, $R_i(0) = 0.0012$, $R_v(0) = 0$, $D_n(0) = 0.0062$, $D_i(0) = 0$, $D_v(0) = 0$, $x_1(0) = 0.2684 N$ and $x_2(0) = 0.1632 N$ where $N = 1000000$.}
\end{figure}

Next, we simulate a scenario with a population already affected by a certain pathogen but with an ongoing vaccination program against the disease. For this simulation, we consider that a vaccine against the disease, already established in the population, has been developed and that a vaccination program has been started afterwards. We then simulated a scenario without a vaccine, as in \ref{infectadosS} and \ref{comportamentosS}, for 500 days and used the information obtained as initial conditions for the scenario with vaccination. See in \ref{infectadosV} and \ref{comportamentosV} the total variation of infected over time and the variation of strong, weak and non-cooperative behaviors in this scenario.

We can observe in \ref{comportamentosV}, that the number of individuals who opt for the strong cooperation strategy decreases considerably in the period from 0 to 2000 days after the start of vaccination and remains very low over time. This can be justified by the fact that the existence of a vaccine brings peace of mind to the population, who now see strong cooperation as unnecessary. The dilemma is between weak cooperation and non-cooperation. Note that while strong cooperation decreases, weak cooperation increases. This can be explained by the persistence of contamination peaks in the population as we can see in \ref{infectadosV}. But over time these peaks decrease and the decrease, even if slow, in the number of infected people in the population contributes to the decrease of weak cooperation and the increase in supporters of non-cooperation, a strategy chosen by the majority of the population. See \ref{comportamentosV}.

\begin{figure}[h]
\begin{center}
	\includegraphics[width= 0.5 \textwidth]{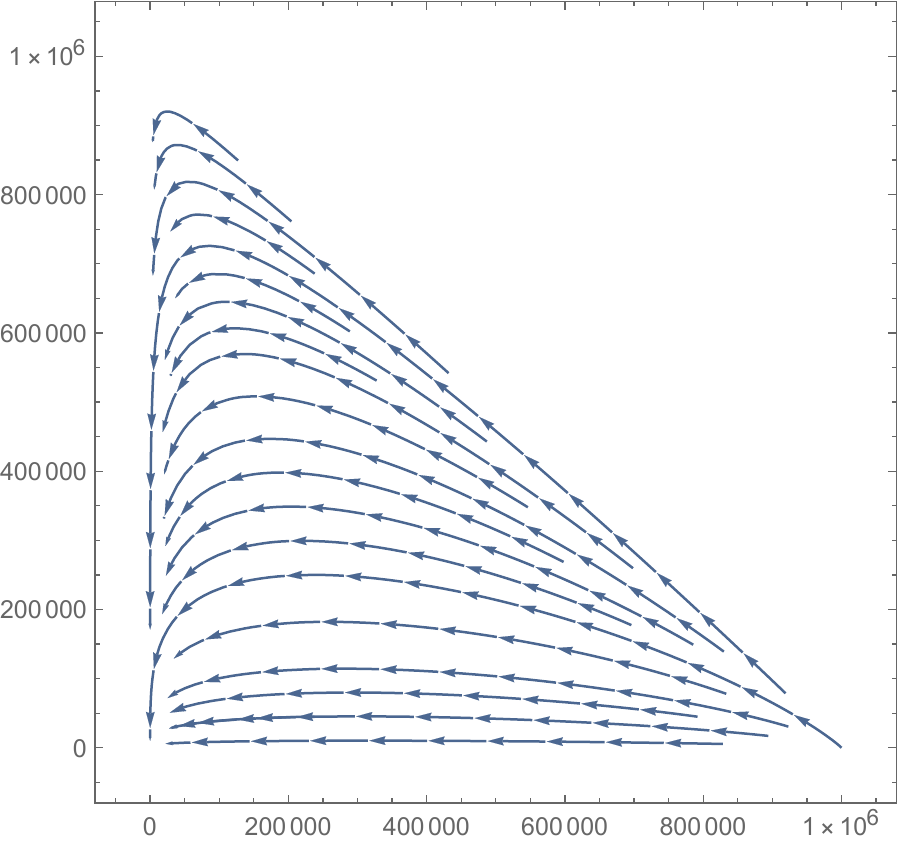}
	\caption{Flux of types of behavior. Close to the equilibrium of the epidemiological model, the number of infected people is close to zero and the variation in deaths is also close to zero, so to simulate the behavior of the orbits of $x_1$ and $x_2$ we use the equations described in \ref{cooperation game sigma} with $G_{a}(I,\dot{D},w) = -a$, $G_{b}(I,\dot{D},w) = -b$ and $G_{c}(I,\dot{D},w) = -c$ and we use the pay-off matrix $\mbox{M} = \left[\begin{array}{ccc} 1 & 0.92 & 0.87 \\ 0.59 & 0.5 & 0.46 \\ 0.1 & 0.03 & 0\end{array}\right]$.}
	\label{fluxoV} 
			\end{center}
\end{figure}

Close to the equilibrium of the epidemiological model, the orbits of the cooperative behaviors considered in this work for the scenario described in \ref{infectadosV} and \ref{comportamentosV} can be seen in \ref{fluxoV}. Note that all orbits converge to the origin, the geometric place that describes the absense of cooperation. What was to be expected from the model as we saw in the previous section on the study of equilibrium points.

\subsection*{Conclusion}
In this work, we proposed an epidemiological model taking into account the natural or induced immunization of individuals. Models that consider the immunization of the population can be very interesting in helping public policies to prevent and contain the disease, since having a forecast of how a disease will spread in the population with an ongoing immunization process can help in planning and choice of the best coping strategy for the disease. The model we present can be used not only in relapsing diseases, for which an immunizer already exists, but also for hitherto unknown diseases, as it considers the immunization acquired when recovering from it. Also, as we have seen happen in the case of COVID-19, there is the possibility of creating an immunizer for a hitherto unknown disease during its dissemination, in which case the model can be applied as described in our study.

The model presented in this work also takes into account the non-linear behavior of the individuals involved. This behavior was modeled using game theory. We used a game with three strategies to model strong cooperation, weak cooperation and non-cooperation behaviors in the population during an epidemic. Coupling the game to the epidemiological model, it was possible to describe and analyze the influence of social behavior on the spread of the disease (looking, for example, at the increase and decrease in the infection rate according to the prevailing social behavior) and also how the scenario epidemiological (the information we have about the disease) can influence our choice of behavior.

\subsection*{Acknowledgments}
The authors would like to thank Fapemig (APQ-01187-22) for its support.


\begin{thebibliography}{10}

\bibitem{amaral2021epidemiological}
M.~A. Amaral, M.~M. de~Oliveira, and M.~A. Javarone.
\newblock An epidemiological model with voluntary quarantine strategies
  governed by evolutionary game dynamics.
\newblock {\em Chaos, Solitons \& Fractals}, 143:110616, 2021.

\bibitem{britton2005essential}
N.~Britton.
\newblock {\em Essential mathematical biology}.
\newblock Springer Science \& Business Media, 2005.

\bibitem{edelstein2005mathematical}
L.~Edelstein-Keshet.
\newblock {\em Mathematical models in biology}.
\newblock SIAM, 2005.

\bibitem{ferreira2020controle}
E.~M. Ferreira, L.~T. Takahashi, and L.~A. D'Afonseca.
\newblock Controle: vacina{\c{c}}{\~a}o e custos no combate a varicela.
\newblock 2020.

\bibitem{gaeta2020simple}
G.~Gaeta.
\newblock A simple sir model with a large set of asymptomatic infectives.
\newblock {\em arXiv preprint arXiv:2003.08720}, 2020.

\bibitem{hofbauersigmund}
J.~Hofbauer and K.~Sigmund.
\newblock {\em Evolutionary Games and Population Dynamics}.
\newblock Cambridge University Press, Cambridge, 1998.

\bibitem{kermack1927contribution}
W.~O. Kermack and A.~G. McKendrick.
\newblock A contribution to the mathematical theory of epidemics.
\newblock {\em Proceedings of the royal society of london. Series A, Containing
  papers of a mathematical and physical character}, 115(772):700--721, 1927.

\bibitem{madeo2022identification}
D.~Madeo and C.~Mocenni.
\newblock Identification and control of game-based epidemic models.
\newblock {\em Games}, 13(1):10, 2022.

\bibitem{MaynardSmithPrice}
J.~{Maynard Smith} and G.~Price.
\newblock The logic of animal conflicts.
\newblock {\em Nature}, 246:15 -- 18, 1973.

\bibitem{murray2007mathematical}
S.~S. .~B. Media, editor.
\newblock {\em Mathematical Biology: I. An Introduction}, volume~17.

\bibitem{morgenstern1953theory}
O.~Morgenstern and J.~Von~Neumann.
\newblock {\em Theory of games and economic behavior}.
\newblock Princeton university press, 1953.

\bibitem{neves2020predicting}
A.~G.~M. Neves and G.~Guerrero.
\newblock Predicando a evolu\c c\~ao da epidemia de covid-19 com o modelo
  a-sir: Lombardia, {I}t\' alia e estado de {S}\~ao {P}aulo, {B}rasil.
\newblock {\em Physica D: Nonlinear Phenomena}, 413:132693, 2020.

\bibitem{smith1982evolution}
J.~M. Smith.
\newblock {\em Evolution and the Theory of Games}.
\newblock Cambridge university press, 1982.

\bibitem{taylorjonker}
P.~D. Taylor and L.~B. Jonker.
\newblock Evolutionary stable strategies and game dynamics.
\newblock {\em Math. Biosci.}, 40:145--156, 1978.

\bibitem{tucker1983mathematics}
A.~W. Tucker and P.~D. Straffin~Jr.
\newblock The mathematics of tucker: A sampler.
\newblock {\em The Two-Year College Mathematics Journal}, 14(3):228--232, 1983.

\end{thebibliography}
\end{document}